\documentclass[12pt]{iopart}

\usepackage{times}
\usepackage{graphicx}
\usepackage{psfig}

\def\be{\begin{eqnarray}}
\def\ee{\end{eqnarray}}

\def\nue{{\nu_e}}

\def\numu{{\nu_{\mu}}}

\def\nutau{{\nu_{\tau}}}

\begin{document}

\title{SNO and the solar neutrino problem}
\author{Sandhya Choubey$^a$\footnote{Speaker}, 
Abhijit Bandyopadhyay$^b$, Srubabati 
Goswami$^c$ and
D.P. Roy$^d$}
\address{
$^a$
Department of Physics and Astronomy, University of Southampton,
Highfield, Southampton S017 1BJ, UK\\
$^b$
Saha Institute of Nuclear Physics, Bidhannagar, Kolkata
700 064, INDIA\\
$c$
Harish-Chandra Research Institute, Chhatnag Road, Jhusi,
Allahabad - 211-019, INDIA
$^d$Tata Institute of Fundamental Research, Homi Bhabha
Road, Mumbai 400 005, INDIA }

\begin{abstract}
We study the implication of the first neutral current (NC) data from SNO.
We perform model independent and model dependent
analyses of the solar neutrino data.
The inclusion of the first SNO NC data
in the model independent
analysis determines the
allowed ranges of $^{8}{B}$ flux normalization
and the $\nu_e$ survival probability more precisely than what was possible from
the SK and SNO CC combination. Transitions to pure sterile states 
are seen to be hugely disfavored however transition to ``mixed'' 
states are still viable with a probability of finding 
about 30\% sterile component in the resultant beam at $1\sigma$. 
We perform global $\nu_e-\nu_{active}$ oscillation analyses
of solar neutrino data including the recent SNO results. 
LMA emerges as a huge favorite while LOW appears at the $3\sigma$ level. 
All the other solutions are disfavored at $3\sigma$ while SMA is 
virtually ruled out. Maximal mixing is disfavored at $3\sigma$. 
We explore in some details the reasons for the incompatibility of the maximal 
mixing solution and the LOW solution with the global solar neutrino data.

\end{abstract}


\section{Introduction}

Solar neutrinos have been detected by radiochemical 
experiments involving capture of
electron type neutrinos by $^{37}Cl$ at Homestake (Cl)
and by $^{71}Ga$ at Sage, Gallex and GNO (Ga)
experiments \cite{Cleveland:nv,globalsolar}. The real time water 
Cerenkov detector Super-Kamiokande (SK) (and earlier the Kamiokande)
have observed the solar neutrinos 
through neutrino-electron scattering \cite{Fukuda:2002pe}. 
All these experiments report a 
deficit of the solar neutrino flux compared to that predicted 
by the Standard Solar Model (SSM) \cite{Bahcall:2000nu}. 
This mystery of missing solar neutrinos consititutes the long standing 
{\it Solar Neutrino Problem}. Neutrino oscillations has been the 
most widely accepted solution for this descrepancy. However in 
the {\it pre-SNO} era, the oscillation hypothesis faced quite a few 
dilemmas. Firstly there was no unambiguous evidence for the 
presence of oscillations from a single experiment. Even though the 
global data suggested that neutrino oscillations may be responsible 
for the depletion of the solar neutrino flux, there seemed to be 
multiple solutions in the neutrino oscillation parameter space -- 
the so called 
Large Mixing Angle (LMA), Small Mixing Angle (SMA), 
LOW mass-squared (LOW), Quasi Vacuum Oscillation (QVO) 
and Vacuum Oscillation (VO) solutions.   
It was also not clear whether 
the $\nue$ converted into another active flavor or disappeared 
into {\it sterile} states. 

The Sudbury Neutrino Observatory (SNO) has now provided for the first
time the direct evidence for oscillations of electron type 
neutrinos to another non-electron type active neutrino, 
en route to earth from
the interior of the sun, by simultaneously observing the charged
current (CC) and neutral current (NC) interactions of neutrinos on 
deuteron \cite{Ahmad:2002jz,Ahmad:2002ka}. 
This single experiment gives evidence for the presence 
of another active neutrino flavor in the solar neutrino beam at the 
$5.3\sigma$ level. When combined with the electron scattering 
data from SK, oscillations to active neutrinos is confirmed at the 
$5.5\sigma$ level. 

In this talk we highlight the impact of the recent SNO results on the 
neutrino oscillation solution to the solar neutrino problem. 
We first study the constraints on the solar neutrino suppression rate 
$P_{ee}$ and the $^8B$ flux normalization factor ($f_B$) 
from SNO and SK in a (quasi)model independent way. 
We derive the $1\sigma(2\sigma)$ limits on $P_{ee}$ and $f_B$ for 
(1) oscillations to only active neutrino states (2) oscillations 
to states that are a mixture of active and sterile components.
For the latter we extract limits on the fraction of the sterile 
component in the solar neutrino beam.

We next inlcude the data from all experiments and 
perform a global statistical analysis in the framework of 
two flavor oscillations to active neutrino states.
The LMA solution is reinstated as the best-fit solution 
but the LOW solution remains allowed at 3$\sigma$ 
\cite{Ahmad:2002ka,Barger:2002iv,Creminelli:2001ij,
Bandyopadhyay:2002xj,Bahcall:2002hv,deHolanda:2002pp,Strumia:2002rv,
Fogli:2002pt,Maltoni:2002ni}.
The SMA solution is comprehensively ruled out while the VO-QVO 
are disfavored at 
$3\sigma$. Maximal mixing is ruled out at more than $3\sigma$.

\section{Model independent bounds on $^8$B flux normalization
and survival probability}

SNO gives us a measure of the total $^8B$ flux coming from the Sun by 
the NC breakup of deuterons. It also gives us the $\nue$ component 
in the $^8B$ beam from the CC reaction on deuterons. Independently, 
SK gives us the ES rate of the $^8B$ neutrinos. The $\nue$ part of 
the $^8B$ flux scatter electrons in SK by the charged current process while 
any other possible active component in the beam ($\numu$ or $\nutau$) would 
scatter electrons through the neutral current channel, however with 
lesser strength. Thus SK has less sensitivity to the total solar 
neutrino flux but it has huge statistics and can be used along with the 
NC data to constrain the total flux and survival probability. Therefore 
as a first step we use only these three pieces of information 
on the observed $^8B$ flux 
in a model independent way, to 
extract maximum information on the total $^8B$ flux produced inside the 
Sun and the rate at which they are suppressed in transit from Sun to 
Earth. 

Both SK and SNO give their data above 5 MeV and both are consistent 
with no energy dependence in the observed suppression rate. 
Hence above 5 MeV we treat $P_{ee}$ to be effectively energy independent 
and express the SK, CC and NC rates as
\be
R^{el}_{SK} &=& f_B P_{ee} + f_B r P_{ea},
\label{rsk} \\[2mm]
R^{CC}_{SNO} &=& f_B P_{ee},
\label{rcc} \\[2mm]
R^{NC}_{SNO} &=& f_B (P_{ee} + P_{ea}),
\label{rnc}
\ee
where $P_{ee}$ denotes the $\nue$ survival probability, 
$P_{ea}$ is the transition probability to active neutrino,
$f_B$ is the $^8B$ normalization factor and  
$r = \sigma_{\nu_\mu,\tau}/\sigma_{\nu_e} \simeq 0.157$ 
is the ratio of $\numu$ to $\nue$ scattering cross-section 
folded with the $^8B$ neutrino spectrum and averaged over energy. 
Note that $r$ depends on the detector characteristics. We have computed 
$r$ for SK above 5 MeV. 
All the rates
are defined with respect to the BPB2000
Standard Solar Model (SSM) \cite{Bahcall:2000nu}.
If we assume that the solar $\nue$ are converted entirely to another 
active flavor then Eq.(\ref{rsk}) and (\ref{rcc}) can be used to 
predict the total observed $^8B$ in SNO 
\cite{Barger:2001zs,Bandyopadhyay:2002bu}
\be
R^{NC}_{SNO} = R^{CC}_{SNO} + (R^{el}_{SK} - R^{CC}_{SNO})/r,
\label{one}
\ee
We showed in \cite{Bandyopadhyay:2002bu} that
because SNO has a greater sensitivity  to the NC scattering rate
as compared to SK, the SNO NC measurement will be
more precise and hence incorporation of this can be more predictive than
the SNO CC and SK combination.
We took three representative NC rates
-- $R_{NC}^{SNO}$ = 0.8,1.0 and 1.2 ($\pm 0.08$)
and showed that
\begin{enumerate}
\item
For a general transition of $\nu_e$ into a mixture of active and sterile
neutrinos the size of the sterile component can be better
constrained than  before.

\item
For transition to a purely active neutrino the $^8{B}$ neutrino flux
normalization and the survival probability $P_{ee}$
are determined more precisely.

\item
We had
also performed global two flavour
oscillation analysis of the solar neutrino
data for the $\nu_e -\nu_{active}$ case, where
instead of $R_{SK}$ and $R^{CC}_{SNO}$ we  used the quantities
$R^{el}_{SK}/R^{NC}_{SNO}$ and $R^{CC}_{SNO}/R^{NC}_{SNO}$.
These ratios are independent of the
$^8B$ flux normalization and hence of the SSM uncertainty.
We showed that use of  these ratios can result in drastic
reduction of the allowed parameter regions specially in the LOW-QVO
area 
depending on the value of the NC rate.

\end{enumerate}
We now have the actual experimental result
\be
R^{NC}_{SNO} = 1.01 \pm 0.12
\ee
while
eq. (\ref{one}) gives $1.05 \pm 0.15$.
Thus in 306 live days (577 days)
the SNO NC measurement has achieved a precision, which is already
better than that obtained from the SK and SNO CC combination.
Since the NC results from SNO rule out transitions to sterile states 
at $5.5\sigma$, we first assume a $\nue-\nu_{\rm active}$ scenario and 
use the actual NC data to derive the limits on $P_{ee}$ and $f_B$. 
We then take a more general approach in which we consider transitions 
to states that are mixture of active and sterile components. We then 
place limits on the sterile admixture in the solar beam using the latest 
data.


\subsection{Case I$:$ Transition of $\nue$ into purely active 
neutrinos}
\begin{figure}
\centerline{\includegraphics[width=9cm]{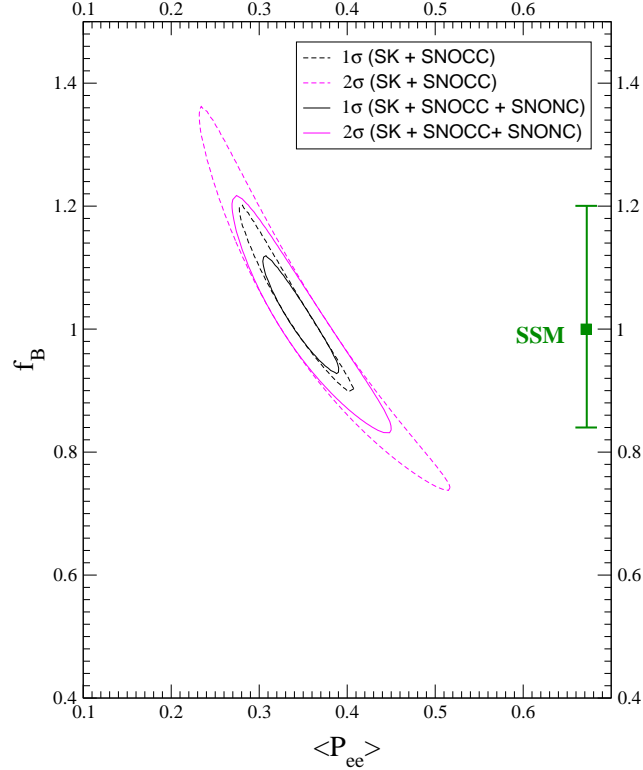}}

\caption{
The $1\sigma$ and $2\sigma$ contours of solutions to the $^8B$
neutrino flux $f_B$ and the $\nu_e$ survival probability $P_{ee}$
assuming $\nu_e$ to $\nu_a$ transition.  The 1$\sigma$ $SSM$ error
bar for $f_B$ is indicated on the right.
}
\label{fig1}
\end{figure}

In this case $P_{ea} = 1 - P_{ee}$ and the equations
(\ref{rsk}), (\ref{rcc}) and (\ref{rnc}) are simplified to 
\be
R^{el}_{SK} &=& f_B P_{ee} + f_B r (1 - P_{ee}),
\label{rsk1} \\[2mm]
R^{CC}_{SNO} &=& f_B P_{ee},
\label{rcc1} \\[2mm]
R^{NC}_{SNO} &=& f_B
\label{rnc1}
\ee
We show in figure
2 the 1$\sigma$ and 2$\sigma$ contours in the $f_B-P_{ee}$
plane from the combinations SK + CC (the outer lines) 
and SK + CC + NC (the inner lines). The best-fit $f_B$ comes at 
1, completely consistent with the SSM.
We note that though 
CC and SK can uniquely determine $P_{ee}$ and 
$f_B$ (cf. Eq.(\ref{one})), the errors are large. This is mainly 
due to the low sensitivity of SK to $f_B$. This error is then carried 
over to $P_{ee}$ due to the strong anticorrelation between $f_B$ and 
$P_{ee}$
through CC. The NC on the other hand is sensitive to $f_B$ within 
almost 10\% and thus 
the inclusion of the NC data narrows down the ranges of $f_B$
and P$_{ee}$.The error in $f_B$ after inclusion of the NC data
is almost half the size of the corresponding error from SSM as is seen 
from figure \ref{fig1}.


\subsection{Case II$:$ Transition of $\nue$ to a mixed state of 
active and sterile 
neutrinos}

We next give up the assumption that the neutrinos transform entirely into 
active states. We instead take up a general case where the $\nue$ 
oscillate into a state $\nu^\prime$ where
\be
\nu^\prime =  \nu_{\rm active} \sin\alpha + \nu_{\rm sterile} \cos\alpha
\label{state}
\ee
$\sin^2\alpha(\cos^2\alpha)$ being the fraction of active(sterile) 
component in the resultant beam on Earth.
In this case $P_{ea} = \sin^2\alpha (1 - P_{ee})$.
Substituting this 
in Eqs. (\ref{rsk}) and (\ref{rnc}) and eleminating $P_{ee}$ using
equation (\ref{rcc}) one gets the following sets of equations for $f_B$ and 
$\sin^2\alpha$
\be
\sin^2\alpha (f_B - R^{CC}_{SNO}) &=& (R^{el}_{SK} - R^{CC}_{SNO})/r,
\label{fbsk}
\\[2mm]
\sin^2\alpha (f_B - R^{CC}_{SNO}) &=& R^{NC}_{SNO} - R^{CC}_{SNO}.
\label{fbnc}
\ee
We treat $\sin^2\alpha$ as a model parameter and for 
different input values of sin$^2\alpha$ we determine the central and
the 1$\sigma$ and 2$\sigma$ ranges of $f_B$ by taking a weighted avarage 
of the equations (\ref{fbsk}) and (\ref{fbnc}). 
The corresponding 
curves are presented in the right-hand panel of figure 2. 
The left-hand panel of the figure shows the curves using 
just the SK and CC data (cf. Eq.(\ref{fbsk})). The NC data is 
seen to put severe restrictions on the allowed values of $\sin^2\alpha$. 
We note that transition to pure 
active components ($\sin^2\alpha=1$) is completely consistent 
with data while transitions to pure sterile states ($\sin^2\alpha=0$) 
are completely forbidden.
Combining the $1\sigma(2\sigma)$ 
lower limit of $f_B$ from this fit with the $1\sigma(2\sigma)$ 
upper limit from SSM (vertical lines are the $2\sigma$ limit)
gives a lower limit of $\sin^2\alpha >0.68(0.45)$.
In other words the sterile fraction in the beam is
\be
\cos^2\alpha <0.32(0.65) ~~{\rm at}~~ 1\sigma(2\sigma)
\label{stbound}
\ee
This means that at $1\sigma(2\sigma)$ we can still have upto
32\%(65\%) admixture of sterile component in the solar neutrino beam. 
There is no upper limit on 
this quantity since the data is perfectly compatible with the $\nue$
transition into purely active neutrinos.

\begin{figure}
\begin{center}
\includegraphics[width=9.0cm]{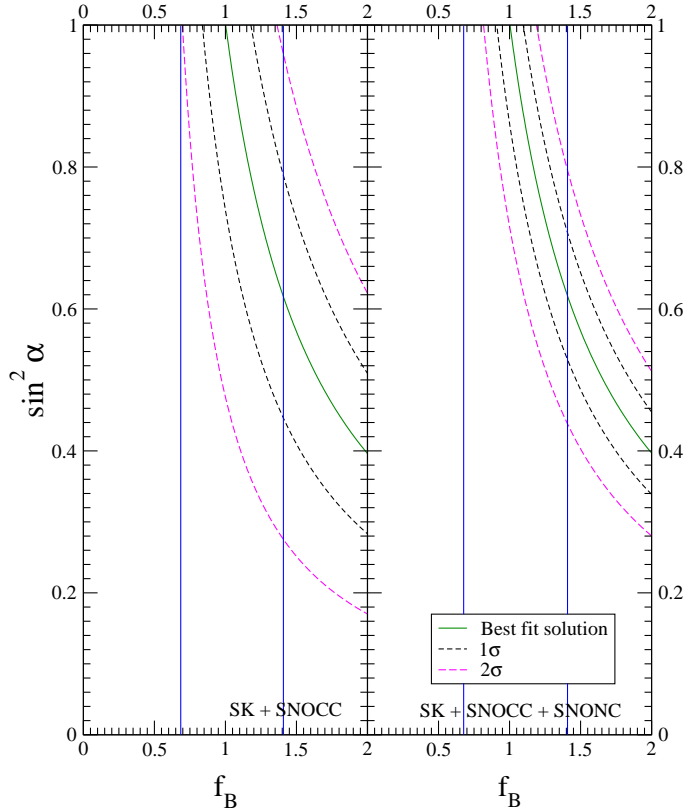}

\caption{
Best fit value of the $^8B$ neutrino flux $f_B$ shown along with
its $1\sigma$ and $2\sigma$ limits against the model parameter
$\sin^2\alpha$,
representing $\nu_e$ transition into a mixed state ($\nu_a \sin\alpha
+ \nu_s \cos\alpha$). The verticle lines denote the $\pm 2\sigma$
limits of the $SSM$. The left-hand panel is for a combination of 
SK+CC. The right-hand panel corresponds to SK+CC+NC data combined. 
}
\end{center}
\label{fig2}
\end{figure}

While we have obtained the bound given by Eq.(\ref{stbound}) in 
a (quasi)model independent way, the same can be obtained in the framework of 
neutrino oscillations \cite{Bahcall:2002zh,Maltoni:2002ni}. 
While \cite{Bahcall:2002zh} puts limits on the sterile component 
within the oscillation hypothesis keeping $^8B$ flux free, the 
authors of \cite{Maltoni:2002ni} fit the data and place bounds 
on $\cos^2\alpha$ with $f_B$ fixed at the SSM value. Our bounds 
(cf. Eq.(\ref{stbound})) is in excellent agreement with the bounds 
obtained in \cite{Bahcall:2002zh,Maltoni:2002ni}.


\section{Model dependent analysis}

In the previous section we restricted ourself to 
the analysis of the observed solar neutrinos rates 
in SK and SNO. We now look at the implications of the 
global solar neutrino data in the framework of two-generation 
oscillations. Since SNO disfavors the sterile option at $5.5\sigma$ 
we consider transitions to active flavors only. For the 
global data we consider the total rates observed in Cl and 
Ga (SAGE, GALLEX and GNO combined rate), the 1496 day SK 
zenith angle energy spectrum data and the recent data from SNO.
Since it is not yet possible to identify the ES, CC and NC events
separately in SNO,
the SNO collaboration have made available their results as a combined
CC+ES+NC data in 17 day and 17 night energy bins.
For the null hypothesis case (which actually corresponds to the case no 
distortion of the solar neutrino energy spectrum) 
they do give the CC and NC (and ES) rates
\cite{Ahmad:2002jz}. These rates would slightly change
with the distortion of the $^8B$ neutrino spectrum from the Sun.
The errors in the CC and NC rates are also 
highly (anti)correlated. However these rates work very well
for studying theories with little or no energy distortion such as the
LMA and LOW MSW solutions if the (anti)correlations between the
CC and NC rates are taken into account. In the previous section we 
used them to study the model independent limits on the 
$^8B$ flux and the suppression rate 
under the assumption 
that there is no energy distortion of the $^8B$ spectrum.
Since we do have an emperical justification to believe that the 
$^8B$ spectrum is indeed undistorted above 5 MeV \cite{Choubey:2001bi}, 
we can use the rates even for an 
oscillation analysis to visualise what impact the NC rate has had 
on the neutrino oscillation parameter space \cite{Bandyopadhyay:2002xj}.
In this section we first present results of a
comprehensive analysis involving the full day-night spectrum of SNO.
We give the best-fit solutions and display the allowed areas in the 
parameter space. We then use CC and NC rates from SNO (instead of the 
day-night spectrum) along with the data from Cl, Ga and SK to 
emphasise the importance of the NC rate.  

\subsection{Global analysis with SNO day-night spectrum}

\begin{figure}
\begin{center}
\includegraphics[width=10.0cm]{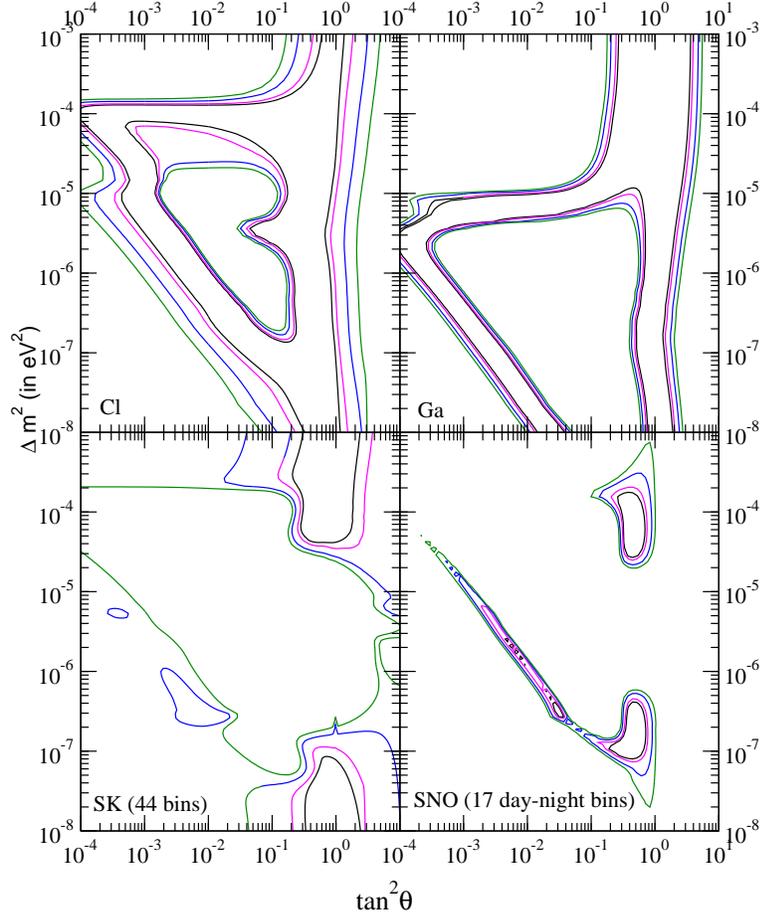}
\caption{The 90\%, 95\%, 99\% and 99.73\% C.L. 
regions of the parameter space allowed by the  
analysis of data from each of the four experiments considered 
separately at a time. We keep $f_B$ fixed at the SSM value. For SK 
we use the zenith angle energy spectra while for SNO we analyse 
the combined SNO day-night spectrum.}
\label{fig3}
\end{center}
\end{figure}
\begin{figure}
\begin{center}
\includegraphics[width=9cm]{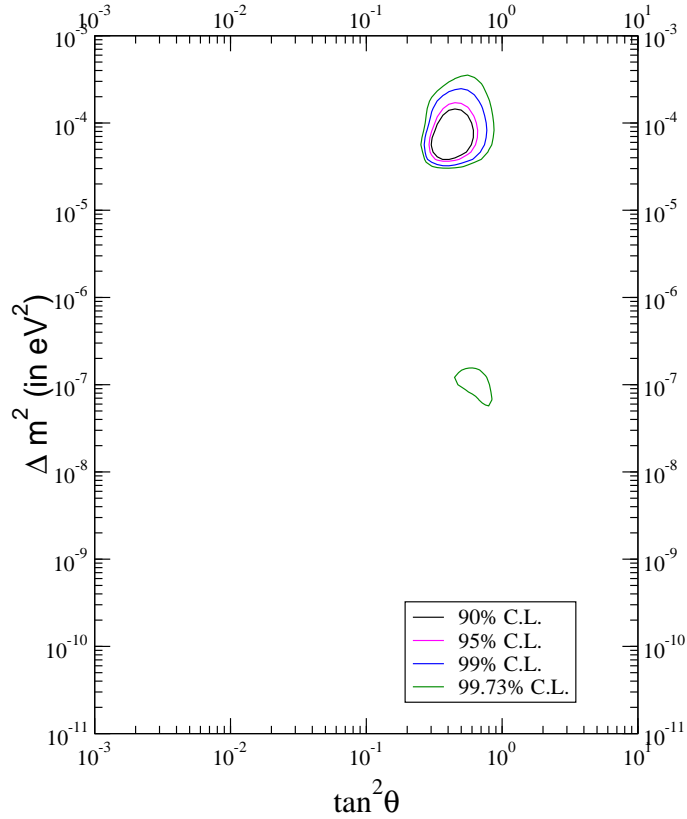}
\caption{The regions in the parameter space allowed from the analysis 
of global solar neutrino data.}
\label{fig4}
\end{center}
\end{figure}

In figure \ref{fig3} we show the allowed areas in the
parameter space from each of the experiments, Cl, Ga, SK and 
SNO. The best-fit for SK comes in the QVO region while SNO 
has its best-fit in the LMA zone. We note that large parts of the 
parameter space are allowed by each of the four experiments.
These allowed zones span LMA as well as LOW-QVO-VO and SMA. 
However only parts of the parameter space which can explain all 
the four experiments simultaneously would be allowed by the 
global data. 
For the global analysis we define the $\chi^2$ function in the 
``covariance'' approach as
\be
\chi^2 = \sum_{i,j=1}^N (R_i^{\rm expt}-R_i^{\rm theory})
(\sigma_{ij}^2)^{-1}(R_j^{\rm expt}-R_j^{\rm theory})
\label{chi1}
\ee
where $N$ is the number of data points ($2+44+34=80$ in our case) and
$(\sigma_{ij}^2)^{-1}$ is the inverse of the covariance matrix,
containing the squares of the correlated and uncorrelated experimental
and theoretical errors. The only correlated error between the total
rates of Cl and Ga, the SK zenith-energy spectrum and the SNO day-night
spectrum data is the theoretical uncertainty in the $^8B$ flux.
However we choose to keep the $^8B$ flux normalization $f_B$ a free
parameter in the theory, to be fixed by the neutral current contribution
to the SNO spectrum. 
We can then block diagonalise the covariance
matrix and write the $\chi^2$ as a sum of $\chi^2$ for the rates,
the SK spectrum and SNO spectrum.
\be
\chi^2 = \chi_{\rm rates}^2 + \chi_{\rm SKspec}^2 + \chi_{\rm SNOspec}^2
\ee
For $\chi^2_{\rm rates}$ we use $R_{\rm Cl}^{\rm expt}=2.56\pm0.23$ SNU
and  $R_{\rm Ga}^{\rm expt}=70.8\pm4.4$ SNU. The details of the theoretical
errors and their correlations that we use can be found in 
\cite{Bandyopadhyay:2001aa}.

For the 44 bin SK zenith angle energy spectra we use the data and
experimental errors given in \cite{Fukuda:2002pe}. SK divides it's
systematic errors into ``uncorrelated'' and ``correlated'' systematic
errors. We take the ``uncorrelated'' systematic errors to be
uncorrelated in energy but fully correlated in zenith angle. The
``correlated'' systematic errors, which are fully correlated in
energy and zenith angle, include the error in the $^8B$ spectrum
shape, the error in the resolution function and the error in the
absolute energy scale. For each set of theoretical
values for $\Delta m^2$ and $\tan^2\theta$ we evaluate
these systematic errors taking into account the relative signs
between the different errors. Finally we take an
overall extra systematic error of $2.75\%$,
fully correlated in all the bins \cite{Fukuda:2002pe}.

For SNO we take the full day-night
spectrum data by adding contributions from CC, ES and NC
and comparing with the experimental results given in \cite{Ahmad:2002jz}.
For the correlated spectrum errors and
the construction of the covariance matrix we follow the
method of ``forward fitting'' of 
the SNO collaboration detailed in \cite{snodata}.

The results of the global $\chi^2$ analysis is shown in Table \ref{table1}.
The best-fit comes in the LMA region as before \cite{Bandyopadhyay:2001aa} but 
LOW is still allowed with a pretty high probability. However SMA is 
seen to be virtually ruled out by the data.
\begin{table}
\begin{center}
\begin{tabular}{cccccc}
\hline
Nature of & $\Delta m^2$ &
$\tan^2\theta$&$\chi^2_{min}$&goodness\\
Solution & in eV$^2$&  &&of fit (\%)\\
\hline
LMA & $6.06 \times 10^{-5}$&$0.41$ & 68.19 &75\% \\
LOW & $1.09\times 10^{-7}$ & 0.59 & 77.49 &46\%\\
VO-QVO& $6.49\times 10^{-10}$& 1.42 & 83.12 &30\% \\
SMA& $4.97\times 10^{-6}$ & $1.57\times 10^{-3}$ & 99.46 &4\%\\
\hline
\end{tabular}
\caption
{The $\chi^2_{min}$, the goodness of fit
and the best-fit values of the oscillation
parameters obtained for the analysis of the global solar neutrino
data including the full SNO day-night spectrum.}
\label{table1}
\end{center}
\end{table}
Figure \ref{fig4} shows the allowed areas in the $\Delta m^2-\tan^2\theta$ 
plane from the global analysis of the solar neutrino data in the 
framework of two-generation $\nue-\nu_{\rm active}$ oscillations. 
Apart from LMA the only other region allowed at $3\sigma$ is the 
LOW solution.
The incorporation of the
recent SNO results narrows down the allowed regions, and in particular
the LOW region becomes much smaller. 
Maximal mixing is seen to be
disallowed at the $3\sigma$ level. The range of values allowed in the 
LMA are
\be
3.8 \times 10^{-5} {\rm eV}^2 \leq \Delta m^2 
\leq 1.4 \times 10^{-4} {\rm eV}^2
\\
0.30 \leq \tan^2\theta \leq 0.62 
~~~~~~~~~~~~~~~~~~~~~~~~~~~~~~~~~~{\rm at}~ 90\% {\rm C.L.}
\ee
\be
3.0 \times 10^{-5} {\rm eV}^2 \leq \Delta m^2 
\leq 3.5 \times 10^{-4} {\rm eV}^2
\\
0.25 \leq \tan^2\theta \leq 0.88 
~~~~~~~~~~~~~~~~~~~~~~~~~~~~~~~~~~{\rm at}~ 
3\sigma
\ee


\subsection{Global analysis with the SNO rates: Impact of the NC data}
\begin{figure}
\begin{center}
\includegraphics[height=12cm,width=12cm]{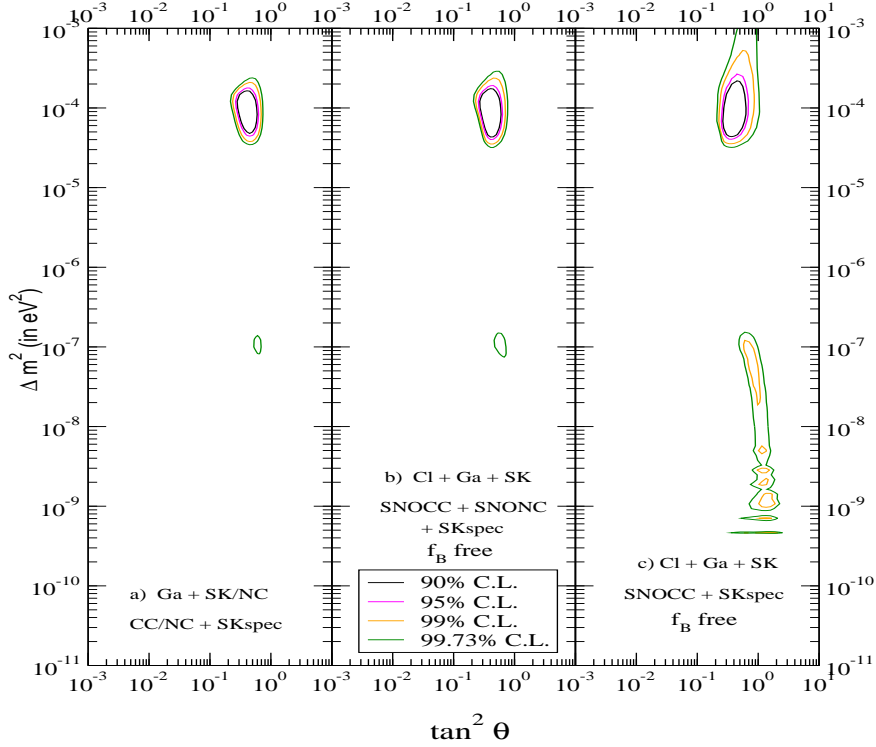}
\caption{The region in the parameter space allowed from the global 
analysis of solar neutrino data including the SNO CC and NC rates 
instead of the SNO spectrum. The panel (c) corresponds to allowed 
area including all data in the analysis except the NC rate.}
\label{fig5}
\end{center}
\end{figure}

To see the impact of the SNO NC results on the oscillation 
solutions 
we replace the SNO day-night spectrum results with the data on 
total CC and NC rates. We use two different approaches for the 
global $\chi^2$ analysis. In the first approach we analyse the global 
data 
using the standard techniques described in our earlier
papers \cite{Bandyopadhyay:2001aa,Goswami:2000wb} except for the fact that
instead of
the quantities  $R^{el}_{SK}$ and $R^{CC}_{SNO}$ we now fit the ratios
$R^{el}_{SK}/R^{NC}_{SNO}$ and
$R^{CC}_{SNO}/R^{NC}_{SNO}$.
The $^{8}{B}$ flux normalization
gets cancelled from these ratios and the analysis becomes
independent of the
large (16-20\%) SSM uncertainty  associated with this.
Since we use both SK rate and SK spectrum data
we keep a free normalization factor for the SK spectrum.
This amounts to taking the information on total rates from the SK rates
data and the information of the spectral shape from the SK spectrum data.
The SNO CC and NC rates have a large anticorrelation. We have
taken into account this correlation between the measured SNO rates
in our global analyses.
Further details of this fitting method
can be found in \cite{Bandyopadhyay:2002bu}.
In Table 2 we present the best-fit parameters and $\chi^2_{min}$.
We have also performed an alternative $\chi^2$ fit to the rates of 
Cl, Ga, SK, CC and NC 
along with the 1496 day SK spectra,
keeping $f_B$ as a free parameter.
Even though we allow $f_B$ to vary freely the NC data serves to control $f_B$
within a range determined by its error. As we see from Table 2
the results of this fit are very similar to the previous cases. The best
fit comes in the LMA region. 

In figure \ref{fig5} we show the allowed area obtained by 
the two different $\chi^2$ analysis procedures with the SNO rates. 
We find that the allowed regions for both approaches using the SNO rates 
are very similar to the ones 
seen in figure \ref{fig4} with the SNO day-night spectrum.
To illustrate the impact of the NC rate
on the oscillation solutions we have repeated the free $f_B$ fit
without this rate. The results are shown in figure \ref{fig5}c.
By comparing figure \ref{fig5}c with figure \ref{fig5}a and 
figure \ref{fig5}b we note that:
\begin{itemize}
\item NC rate disfavors the LOW solution, which reduces in size and 
appears only at $3\sigma$. The area around LOW-QVO 
need low values of $f_B$ to explain the global rates. However NC 
does not allow such low values of $f_B$ anymore.
\item Values of $\Delta m^2$ above $3.5\times 10^{-4}$ eV$^2$ are
disfavored because these regions need low $f_B$ to remain allowed 
which is not possible with the NC rate.
\item Maximal Mixing is disfavored at $3\sigma$ again because 
NC severely constraints $f_B$.
\item SMA is further disfavored because there is a huge tension 
between the data from Cl+Ga and SK+SNO.
\item Dark Side solutions are gone, which implies that 
$\Delta m^2_{\rm solar} > 0$.
\end{itemize}
\begin{table}
\begin{center}
\begin{tabular}{cccccc}
\hline
Data&Nature of & $\Delta m^2$ &
$\tan^2\theta$&$\chi^2_{min}$& Goodness\\
Used&Solution & in eV$^2$&  & & of fit\\
\hline
Ga + &LMA & $9.66 \times 10^{-5}$&$0.41$ & 35.95 & 80\%\\
SK/NC + &LOW & $1.04\times 10^{-7}$ & 0.61 & 46.73 & 36\%\\
CC/NC + &VO-QVO& $4.48\times 10^{-10}$& 0.99 & 54.25  & 14\%\\
SKspec&SMA& $6.66\times 10^{-6}$ & $1.35\times 10^{-3}$ & 67.06 & 1\%\\
\hline
Cl + Ga +&LMA & $6.07 \times 10^{-5}$&$0.41$ & 40.57 & 66\%\\
SK + CC +&LOW & $1.02\times 10^{-7}$ & 0.60 & 50.62 & 26\%\\
NC + SKspec &VO-QVO& $4.43\times 10^{-10}$& 1.1 & 56.11  & 12\%\\
+ $f_B$ free&SMA& $5.05\times 10^{-6}$ & $1.68\times 10^{-3}$ & 70.97 & 1\%\\\hline
\end{tabular}
\caption
{The $\chi^2_{min}$, the goodness of fit
and the best-fit values of the oscillation
parameters obtained for the analysis of the global solar neutrino
data. The data set used is shown.}
\end{center}
\end{table}

\section{Any chance for maximal mixing?}

\begin{figure}[b]
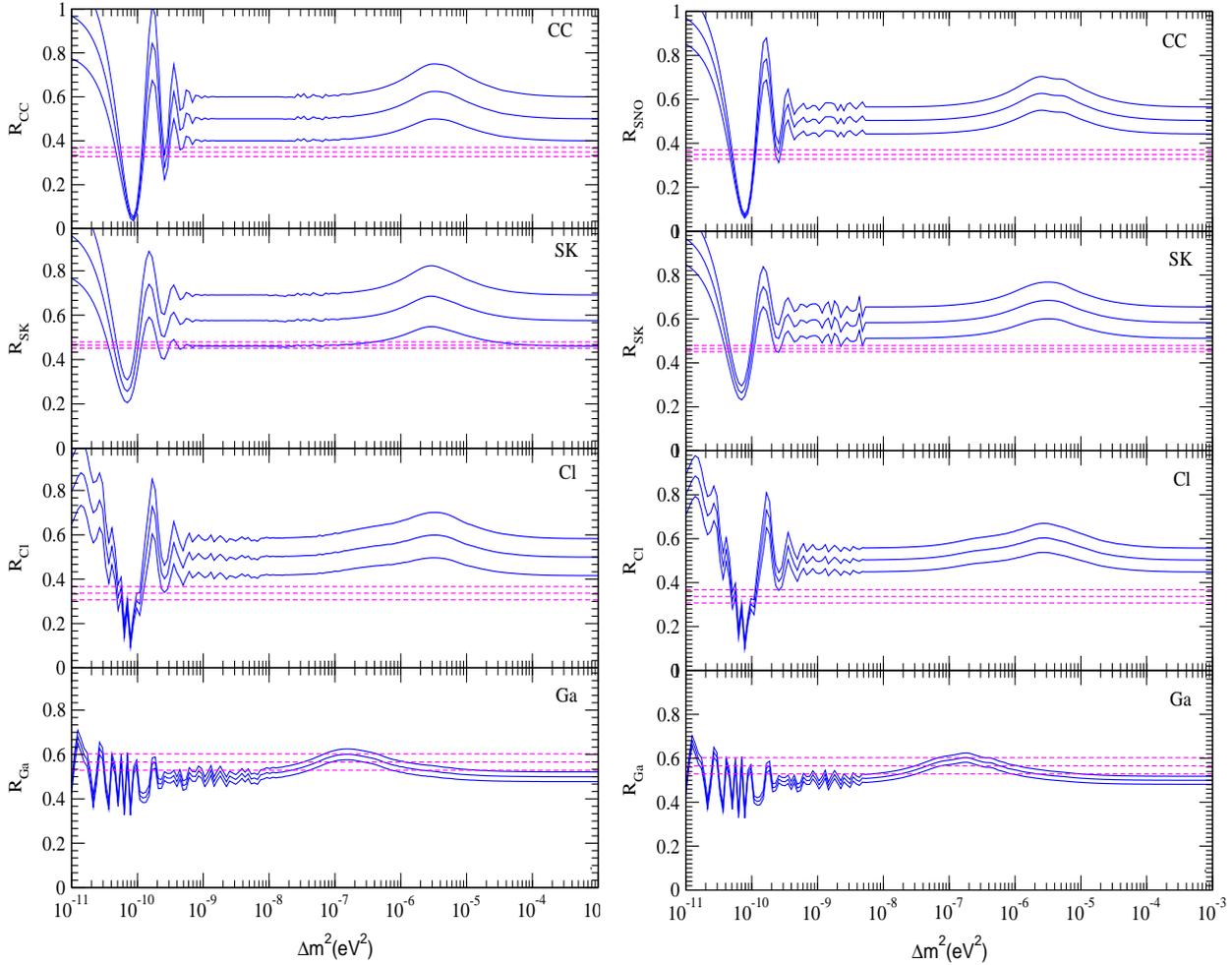

\begin{center}
\vglue 0.0cm \hglue -8.5cm
\includegraphics[height=13cm,width=8.2cm]{max1.eps}
\vglue -13.0cm \hglue 7.8cm
\includegraphics[height=13cm,width=8.2cm]{max2.eps}
\caption{
The solar neutrino rates predicted for maximal mixing for SNO CC, SK, 
Cl and Ga, as a function of $\Delta m^2$. The solid lines in the 
left-hand panel show the predicted band for $\pm 1\sigma$ uncertainties 
coming from the theoretical uncertainties for the predicted 
fluxes in BPB00 \cite{Bahcall:2000nu}. The right-hand panel shows the 
corresponding bands of predicted values where the 
theoretical $\pm 1\sigma$ uncertainties in the $^8B$ flux is replaced by 
the error in the NC rate at SNO. The dashed lines show the $\pm 1\sigma$ 
band of the rates observed in these experiments.
}
\label{fig6}
\end{center}
\end{figure}

We next explore in some detail the reason for maximal mixing getting 
disfavored. For maximal mixing the the most general expression 
for the survival probability of the solar 
neutrinos is
\be
P_{ee} = \frac{1}{2} + f_{reg}
\ee
where $f_{reg}$ is the Earth regeneration factor. Clearly $P_{ee}$ is 
never less than 1/2, irrespective of the value of $\Delta m^2$ and neutrino 
energy, and is greater than 1/2 for values of $\Delta m^2$ and neutrino 
energy where Earth regeneration effects are significant. In figure 
\ref{fig6} we show as a function of $\Delta m^2$  
the rates predicted in the four solar neutrino 
experiments at maximal mixing. The solid lines in the 
left-hand panel give the $\pm 1\sigma$ predicted bands for the rates 
taking the $\pm 1\sigma$ error in the solar fluxes 
from BPB00 \cite{Bahcall:2000nu}. The right-hand panel gives the 
corresponding bands when the SSM error in the $^8B$ flux is replaced 
by the experimental error in SNO NC. It is seen that maximal mixing 
can only explain the Ga rate well, since Ga is the only observed rate that 
is greater than 1/2. The reason being the Earth regeneration effects 
for the low energy neutrinos with $\Delta m^2$ in the LOW region. 
In the pre-SNO era since the error in $^8B$ flux was more, even SK is 
seen to be consistent within $\pm 1\sigma$ of the predicted rate, though 
SNO CC and Cl are seen to 
be inconsistent with maximal mixing 
even with SSM uncertainty on $f_B$. Thus because both 
Ga and SK were explained by maximal mixing in the LOW region, it was 
allowed before the advent of SNO NC data \cite{Choubey:2001bi}. 
However with the recent SNO constraints on $f_B$,
we find from the right panel of figure \ref{fig6} that none of the 
experiments except Ga remain consistent with maximal mixing. Thus 
maximal mixing is ruled out by the global data at more than $3\sigma$.

\section{Any chance for LOW?}

\begin{figure}
\begin{center}
\includegraphics[width=13cm]{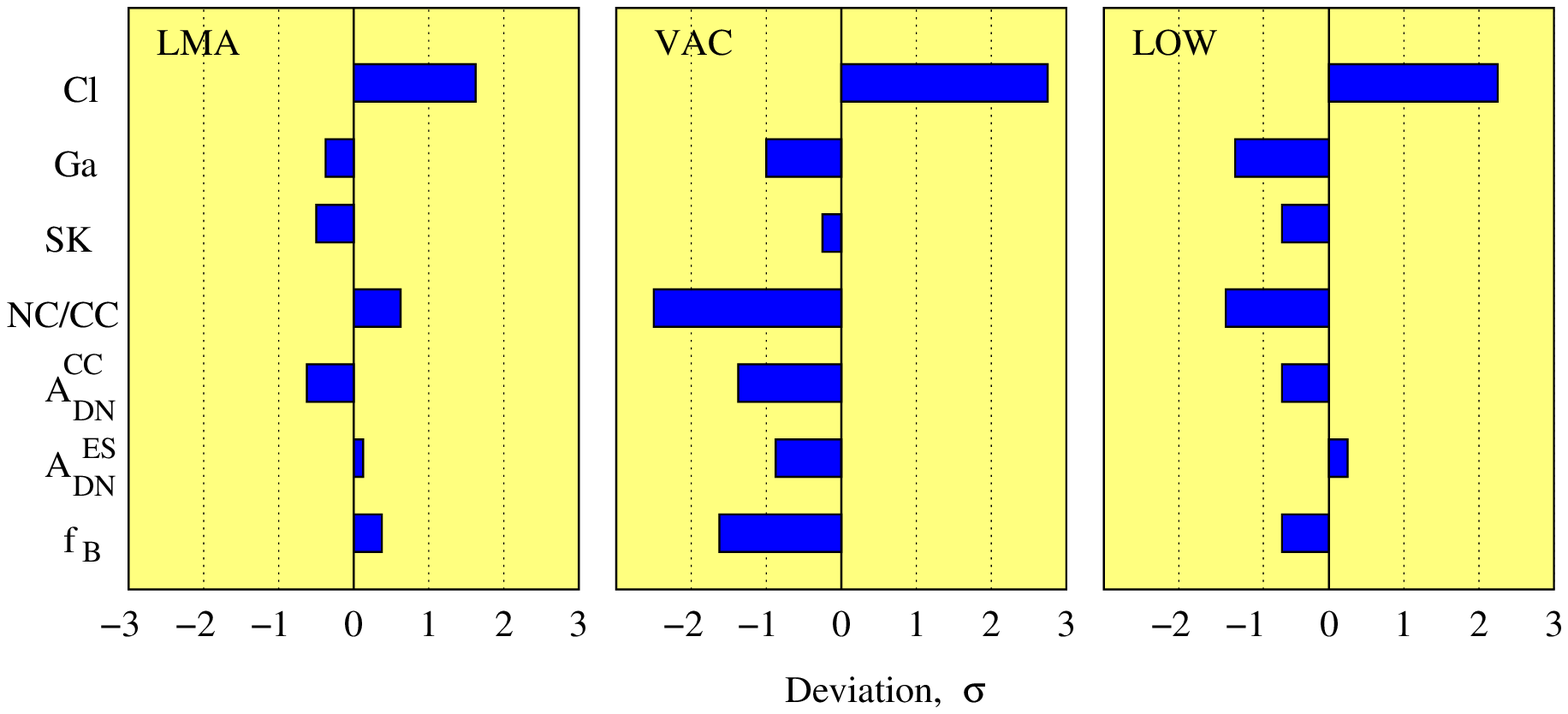}
\vskip -10.5cm
\caption{
Pull-off diagrams for the LMA, VO(VAC) and LOW solutions. 
Shown are the deviations of the predicted values from the 
experimental observations in units of $1\sigma$ of the 
experimental error. This figure is from \cite{deHolanda:2002pp}.

}
\label{fig7}
\end{center}
\end{figure}

From the figures \ref{fig4} and \ref{fig5} it is very clear that  
the LOW solution has fallen into disfavor with SNO. The figure 
\ref{fig7} (taken from \cite{deHolanda:2002pp}) 
shows the ``pulls'' for all the observables in LMA, VO(VAC) and LOW. It 
shows the deviation of the theoretical prediction from the experimental 
results in units of the $1\sigma$ experimental errors. 
The figure clearly shows that while LMA is consistent within $1\sigma$ 
with all the observables except Cl, LOW has some inconsistency. 
In particular, LOW is seen to be 
inconsistent with Ga and SNO, in addition to Cl. 
In ref.\cite{Fogli:2002pt} the authors have made a rigorous global 
analysis using the ``pull method''. We refer the reader to 
\cite{Fogli:2002pt} for the details of the pulls in the observables and 
the systematics. What we focus here is on the 
qualitative understanding of the problem LOW faces in simultaneously 
fitting the Ga and SK-SNO data. 

In the LOW region transitions inside the Sun are almost adiabatic 
(except for the very low $\Delta m^2$). Also since for these 
$\Delta m^2$ the resonance density  
for neutrinos inside the Sun is very much smaller 
(for almost all neutrino energies) 
compared to the density at which they are produced, the mixing angle at 
their point of production is close to $\pi/2$ for all neutrinos. In other 
words, in the LOW regime the survival probability for the neutrinos 
at the surface of the Sun 
has almost no energy dependence. The only energy dependence in the resultant 
survival probability at the detector comes from the Earth regeneration 
effects, which are significant for the low energy neutrinos in LOW.
Therefore the expected rates in Ga (which predominantly observes the $pp$ 
neutrinos) and SK/SNO (which observe the $^8B$ neutrinos) can be written as 
\be
R_{Ga} \approx \sin^2\theta + f_{reg}
\label{ratelowga}
\\
R_{SK} \approx R_{CC} \approx f_B\sin^2\theta
\label{ratelowsk}
\ee 
Thus LOW can explain the high Ga rate simulataneously with the lower 
SK and CC rates only if the Earth regeneration effects are large 
and/or 
$f_B$ is lower than 1. However $f_B$ is constrained to be 
close to 1 now by the NC observations in SNO. This leaves just 
$f_{reg}$ to simultaneously explain $R_{Ga}$ and $R_{SK}/R_{CC}$. 
But again larger values of $\Delta m^2$ for which we have large 
$f_{reg}$ in Ga run into problem with the SK zenith angle energy spectra.
The higher $\Delta m^2$ ($>10^{-7}$ eV$^2$) predict strong peaks in the 
zenith angle spectrum at SK \cite {Gonzalez-Garcia:2000dj}. However 
the zenith angle data at SK is consistent with flat and this rules out 
the higher end of the LOW solution where the Earth regeneration is 
significant. Thus the LOW solution cannot reconcile the high rate 
observed in Ga with the rates seen in SK and SNO CC because 
(i)$f_B$ is constrained by SNO NC and (ii) $f_{reg}$ has to be small 
to be consistent with SK zenith angle energy spectrum.

\begin{figure}[t]
\vglue 0.0cm \hglue -0.8cm
\includegraphics[width=8.5cm]{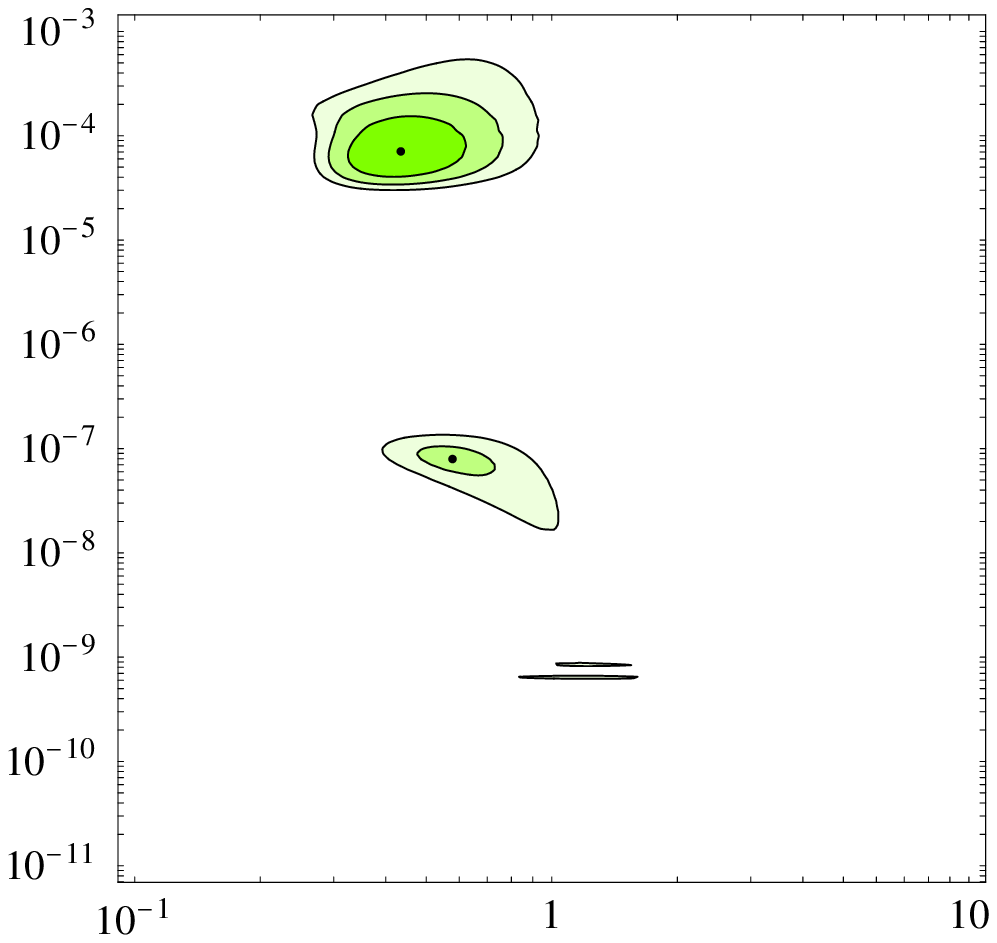}

\vglue -8.5cm \hglue 7.8cm
\includegraphics[width=8.5cm]{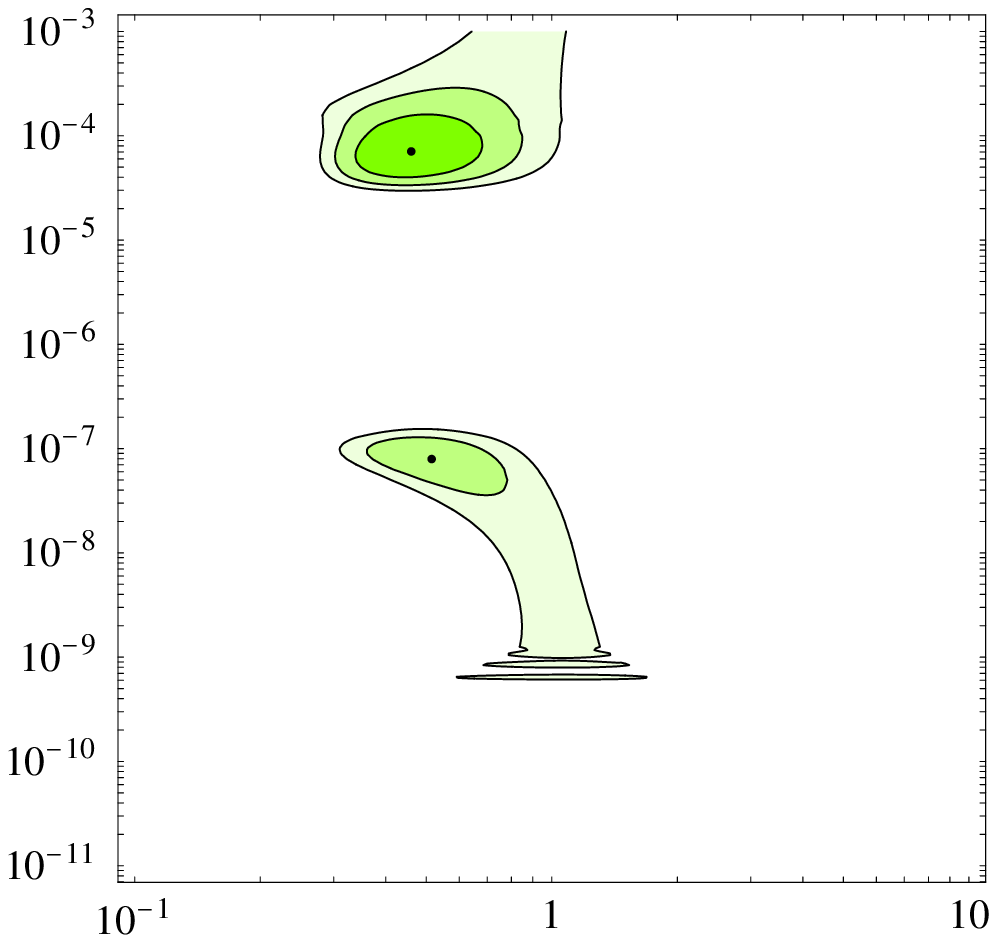}
\vskip -0.5cm
\caption{
The comparison of the area allowed in the neutrino parameter space 
for different values of the Ga rate. The left-hand panel shows 
the 90\%, 99\% and 99.9\% C.L.
areas allowed using all Gallium data ($70.8\pm 4.4$ SNU) while 
the right-hand panel displays the improvement in LOW when only the 
most recent Gallium data ($66.1\pm 5.3$ SNU) is taken. 
This is same as figure 2  
of \cite{Strumia:2002rv} but with the full SNO day-night spectrum 
\cite{strumia}.
}
\label{fig8}
\end{figure}

If Ga is left out of the fit then the difference between the 
LMA and LOW $\chi^2$ comes down to $\Delta \chi^2\approx 3$ 
\cite{Strumia:2002rv}. LOW also improves if the Ga rate was lower.
In fact the observed rates in both SAGE and GALLEX-GNO have been 
lower in recent times. If the data are divided into two periods 
with $1^{\rm st}$ period before April 1998 and the $2^{\rm nd}$ 
period after April 1998 then the combined Ga rates are \cite{Strumia:2002rv}
\be
{R_{Ga}~ (1^{\rm st} period}) &=& 76.4\pm 5.4 ~{\rm SNU}\\
{R_{Ga}~ (2^{\rm nd} period}) &=& 66.1\pm 5.3 ~{\rm SNU}
\ee
The figure \ref{fig8} (from \cite{Strumia:2002rv} but for the full 
SNO day-night spectrum) shows the 
allowed area obtained by the authors of \cite{Strumia:2002rv} from 
a global analysis of the solar data. The left-hand panel shows the area 
obtained using the combined Ga rate ($70.8\pm4.4$ SNU) while the 
right-hand panel shows how the LOW solution improves if the Ga data 
for only the $2^{\rm nd}$ period ($66.1\pm 5.3$ SNU) is included.

\section{Conclusions}

The SNO experiment for the first time gives unambiguos signal 
for conversions (oscillations) of the solar $^8B$ neutrinos 
into a different active neutrino flavor. The comparison of the 
charged current rate with the neutral current data at SNO confirms 
the presence of the ``other'' flavor in the solar neutrino beam at 
the $5.3\sigma$ level. Combined with the electron scattering data 
from SK the recent results from SNO rule out transitions to sterile 
states at the $5.5\sigma$ level. We analysed the recent SNO results 
along with the results from the SK experiment in a model independent way.
We put limits on the allowed ranges for the $^8B$ flux and the neutrino 
suppression rate. We extend our analysis to include transition to 
states which are mixtures of active and sterile components. We find that 
even though transitions to pure sterile states are comprehensively 
ruled out by SNO, transitions to ``mixed'' states are not and the 
resultant solar neutrino beam may have as much as 32\%(65\%) of 
sterile admixture at $1\sigma(2\sigma)$. This bound is consistent 
with the limits obtained from analysis of the global data in the 
framework of neutrino oscillations.

We next include the global solar neutrino data and perform a 
statistical analysis with the ``covariance'' approach under 
the hypothesis of two-generation oscillations 
involving active neutrinos. We first include the full SNO day-night spectrum 
along with the data on total rates from Cl and Ga, and the SK 
zenith angle energy spectra, and present the solutions and the allowed 
regions in the neutrino oscillation parameter space.
Next in order to take a closer look at the impact of the NC data 
on the global solutions, we replace the SNO day-night spectrum with 
the data on CC and NC total rates. We use the flat SK spectrum as a 
justification for using the CC and NC SNO rates, which have been extracted 
for no spectral distortion of the $^8B$ neutrinos, and incorporate 
the (anti)correlation between them.
We find that the SNO neutral current data favors the LMA solution.
The LOW solution even though still allowed by the global data is 
less favored compared to the LMA solution and appears only at 
the $3\sigma$ level. However SMA solutions gets hugely disfavored 
and is virtually ruled out. QVO and VO are also disfavored at more than 
$3\sigma$. Maximal mixing, which was allowed in the LOW region 
prior to SNO -- thanks to the Ga rate and big uncertainty in $f_B$ -- is 
disfavored by the recent SNO results by more than $3\sigma$.

Thus LMA survives as the only strong solution to the solar neutrino 
problem with $3.0 \times 10^{-5}$ eV$^2 \leq \Delta m^2 
\leq 3.5 \times 10^{-4}$ eV$^2$ and $0.25 \leq \tan^2\theta \leq 0.88$ 
at $3\sigma$. The KamLAND reactor (anti)neutrino experiment,
which has unprecedented sensitivity over the entire LMA zone, will 
very soon confirm or refute the LMA solution to the solar neutrino 
problem. If KamLAND does not observe a positive signal 
for oscillations then we will have to wait untill Borexino starts 
taking data to test LOW. The LOW solution, if correct, should give 
a big day-night asymmetry in Borexino. Hopefully we do not have to wait 
long to see the long standing solar neutrino problem completely resolved.

\section*{Acknowledgments}
S.C. would like to thank H. V. Klapdor-Kleingrothaus,
J. Peltoniemi and all other organizers of Beyond the Desert, 2002
for their hospitality during the conference.



\end{document}